\documentstyle[12pt]{article}
\begin{document}
\bibliographystyle{unsrt}
\newcommand{\halfthin}{\kern 0.0834em}
\newcommand{\bra}[1]{\left < \halfthin #1 \right |\halfthin}
\newcommand{\ket}[1]{\left | \halfthin #1 \halfthin \right >}

\title{
$I=0,1$ $\pi\pi$ and $I=1/2$ $K\pi$ Scattering\\
using Quark Born Diagrams\\}

\vskip 0.5cm
\author{Z.Li$^1$,
M.Guidry$^2$,
T.Barnes$^2$ and E.S.Swanson$^3$\\
\\
$^1$Physics Department, Carnegie-Mellon University\\
Pittsburgh, PA 15213\\
\\
$^2$Physics Division and\\
Center for Computationally Intensive Physics\\
Oak Ridge National Laboratory\\
Oak Ridge, TN 37891\\
and\\
Department of Physics and Astronomy\\
University of Tennessee\\
Knoxville, TN 37996\\
\\
$^3$Center for Theoretical Physics\\
Laboratory of Nuclear Science and Department of Physics\\
Massachusetts Institute of Technology\\
Cambridge, MA 02139}
\date{}
\maketitle

\begin{abstract}
\baselineskip=24pt
We extend the quark Born diagram formalism for hadron-hadron scattering to
processes with valence $q\bar q$ annihilation, specifically $I=0$ and $I=1$
$\pi\pi$ and $I=1/2$ $K\pi$ elastic scattering. This involves the $s$-channel
hybrid annihilation process $q^2\bar q^2 \to q\bar q g \to q^2\bar q^2$ and
conventional $s$-channel $q\bar q$ resonances, in addition to the $t$-channel
gluon exchange treated previously using this formalism. The strength of the
$t$-channel gluon amplitude is fixed by previous studies of $I=2$ $\pi\pi$ and
$I=3/2$ $K\pi$ scattering. The $s$-channel resonances $\rho(770)$, $K^*(892)$,
$f_0(1400)$ and $K_0^*(1430)$ are incorporated as relativized Breit-Wigner
amplitudes, with masses and energy-dependent widths fitted to experimental
phase shifts. The strength of the $s$-channel gluon ``hybrid" annihilation
diagrams is problematical since the perturbative massless-gluon energy
denominator must be modified to account for the effect of confinement on the
energy of the virtual hybrid state. Our naive expectation is that near
threshold the hybrid diagrams are comparable in magnitude but opposite in sign
to the contribution predicted using massless perturbative gluons. Fitting the
strength of this amplitude to $I=0$ $\pi\pi$ and $I=1/2$ $K\pi$ S-wave data
gives a result consistent with this expectation. We find good agreement with
experimental phase shifts from threshold to 0.9 GeV in $\pi\pi$ and 1.6 GeV in
$K\pi$ using this approach. We conclude that the most important contribution to
low energy S-wave scattering in these channels arises from the nonresonant
quark Born diagrams, but that the low-energy wings of the broad $s$-channel
resonances $f_0(1400)$ and $K^*_0(1430)$ also give important contributions near
threshold. The nonresonant contributions are found to be much smaller in $L\;
>0$ partial waves, but may be observable in $I=1$ $\pi\pi$ and $I=1/2$ $K\pi$
P-wave phase shifts.
\end{abstract}

\newpage
\subsection*{\bf 1. Introduction}
The determination of
hadron-hadron interactions in terms of quark and gluon
degrees of freedom is an important goal of the
nonrelativistic quark potential model.
Historically these studies first
concentrated
on the nucleon-nucleon interaction, which is well determined
experimentally
and is fundamental to nuclear physics.
Nonperturbative techniques such as the resonating group method \cite{NNRG}
and variational approaches \cite{MI} lead to reasonable descriptions of the
short- and intermediate-range
nucleon-nucleon interaction in terms of quark forces. (For a review of
this work on $NN$ to 1989 see Shimizu \cite{NNrev}.)
Studies of other hadron-hadron channels
in terms of quark and gluon interactions have also been successful.
Weinstein and Isgur \cite{WI}
carried out
variational
studies of
pseudoscalar meson-meson interactions
using the nonrelativistic quark potential model, and found weakly-bound
deuteron-like
``${K\bar  K}$
molecules", which
have been identified with the $f_0(975)$ and $a_0(980)$.
They also used their variational techniques to extract equivalent
low-energy meson-meson potentials, which with some qualifications give
reasonable results for
$\pi\pi$ and $K\pi$ elastic phase shifts \cite{WIKP,WKPS}.
The resonating group techniques have also been applied to the kaon-nucleon
system \cite{KNRG} and give results similar to the observed
$I=0$ and $I=1$ S-wave phase shifts. There may be some discrepancies, however,
and the
$I=0$ $KN$ phase shift in particular is not yet well determined experimentally
and merits further study, which may be possible at
DAPHNE \cite{DAPHNE}.

Recently Barnes and Swanson \cite{BS1,S1} introduced a perturbative
Born-order formalism
for hadron-hadron scattering in terms of quark and gluon degrees of freedom,
based on a single interaction, usually one-gluon-exchange
(OGE) followed
by quark line interchange.
This
analytical technique
reproduces many of the successes of the
earlier nonperturbative calculations when
applied to scattering
processes which are free of
valence quark annihilation. In some cases such as
the $I=2$ $\pi\pi$ and $I=3/2$ $K\pi$ S-waves the Born diagrams
are
in remarkably good agreement with experiment
from threshold to the maximum
experimental invariant mass of about
1.5 GeV.
The reactions studied to date using this method are
$I=2$ $\pi\pi$ \cite{BS1,S1}, $I=3/2$ $K\pi$
\cite{BSW}, certain vector-vector channels
\cite{S1,DSB}, $I=0,1$ $KN$ \cite{BS92} and $NN$, $N\Delta$ and
$\Delta\Delta$ \cite{BCKS}, and several other related channels.
One conclusion of these studies was that
the
powerful but complicated
nonperturbative techniques were unnecessary in some channels,
notably $\pi\pi$ and $K\pi$, because the perturbative
amplitude alone gives a good description of the data.

Motivated by the successes of this simple perturbative approach,
in this paper we generalize this technique to scattering
processes with valence annihilation.
General hadron-hadron scattering amplitudes
include contributions from
resonance production, resonance exchange and
$q\bar q$ annihilation
($s$-channel gluon exchange), in addition to the $t$-channel gluon exchange
considered previously in this formalism.
In principle each of these contributions may
be important, so reactions in annihilation channels can be quite
complicated. Here we consider pseudoscalar-pseudoscalar (Ps-Ps)
elastic scattering with valence annihilation,
and
assume that the important
amplitudes arise from $t$-channel gluon exchange,
$s$-channel gluon exchange  and
$s$-channel resonance production.
The first contribution is treated using the quark Born diagram techniques
developed previously, and we find that the $t$-channel gluons
which dominated $I=2$ $\pi\pi$ and $I=3/2$ $K\pi$ are
numerically
rather weak here.
The second scattering mechanism,
$s$-channel gluon exchange, is treated using the same quark Born
techniques.
Special
care must be taken with this process, however, to incorporate the
mass of the intermediate hybrid state. This changes $s$-channel gluon
exchange
from a repulsive to an attractive interaction
in the cases studied here, and makes the overall magnitude
of this effect rather uncertain. Here we fit it to
experimental S-wave phase shifts,
and compare the fitted strength to our naive estimate based on an
effective gluon mass of $\approx 1$ GeV.
The third contribution, from
$s$-channel resonances, is incorporated
phenomenologically by treating these as relativized Breit-Wigner
phase shifts with free masses and widths, which we fit to the data.

In section 2
we describe our techniques in detail for $I=0$ and $I=1$ $\pi\pi$
scattering, and give numerical estimates of the contribution of each
effect to the scattering length.
In section 3
we give the corresponding results for $I=1/2$ $K\pi$ scattering, which
is formally similar
to $\pi\pi$
but is somewhat more complicated due to the strange
quark mass. In section 4 we carry out detailed fits to experimental data
sets for $I=0$ and $I=1$ $\pi\pi$ and $I=1/2$ $K\pi$
phase shifts and discuss the
relative importance of resonant and nonresonant contributions.
We conclude with a brief summary of our results and
suggestions for future work.

\subsection*{2. Detailed Results for $I=0$ and $I=1$ $\pi\pi$
Scattering Amplitudes}

We consider three contributions to hadron-hadron
scattering in annihilation channels,
\vskip 0.5cm

\noindent
1) $t$-channel gluon exchange,

\noindent
2) $s$-channel gluon exchange (valence $q\bar q$ annihilation), and

\noindent
3) $s$-channel resonances.

\vskip 0.5cm
\noindent
A fourth contribution which is often discussed is $t$-channel meson exchange.
We exclude this scattering mechanism
in $\pi\pi$ and $K\pi$ scattering because the most important such
contribution, one pion exchange, is not present in Ps-Ps scattering, and
in any case
the contribution of $t$-channel
meson exchange to low-energy hadron-hadron scattering has probably been
overestimated. For a discussion of this issue see \cite{NIrev}.

Taking the three mechanisms in order,
their contributions to the scattering amplitude are
as follows:
\vskip .5cm
\noindent
1) {\it t-channel gluon exchange}
\vskip .5cm

In earlier studies of
Ps-Ps
elastic scattering
[4-6] it was found that the dominant scattering mechanism
involves the OGE spin-spin hyperfine interaction,
which between quarks $i$ and $j$ is
\begin{equation}\label{1}
H_{ij} = \;
-{8\pi\alpha_s \over 3 m_i m_j}\; (\lambda^a_i/2)\cdot (\lambda^a_j/2)\;
 (\vec S_i \cdot \vec S_j)\; \delta (\vec r_{ij}) \ .
\end{equation}
A similar conclusion was reached earlier for $NN$ scattering
[1-3]. An explanation of hyperfine dominance in
Ps-Ps scattering was given by Barnes and Swanson \cite{BS1,S1},
who found that the
matrix elements of
$\lambda^a_i \cdot \lambda^a_j\;
\vec S_i\cdot \vec S_j$
in the four Born-order quark
scattering diagrams for
$(q\bar q)(q\bar q)\to
(q\bar q)(q\bar q)$
all have
the same sign in this channel. In contrast, the spin-independent
matrix elements
of
$\lambda^a_i \cdot \lambda^a_j$
(which multiply the spin-independent color Coulomb and linear confining
terms)
interfere destructively between the diagrams.
In the Ps-Ps annihilation channels we consider here the same conclusions
apply,
so we again assume dominance of the
spin-spin color hyperfine term
in $t$-channel gluon exchange.

The $t$-channel gluon exchange contribution to $I=0$ $\pi\pi$ scattering
can be determined from the previous quark Born study of
$I=2$ $\pi\pi$ scattering \cite{BS1}. The only difference
is the flavor factor,
which gives a Hamiltonian matrix element of

\begin{displaymath}
h_{fi}^{I=0}(t{\rm-ch.}\ {\rm gluon}) = -\frac 12
h_{fi}^{I=2}(t{\rm-ch.}\ {\rm gluon}) =
\end{displaymath}
\begin{equation}
-\frac {4\pi\alpha_s}{9m_q^2}\frac 1{(2\pi)^3} \left[
\exp\bigg\{- {(1-\mu)} \frac {k^2}{4\beta^2_\pi}\bigg\}
+\exp\bigg\{- {(1+\mu)}\frac {k^2}{4\beta^2_\pi}\bigg\}
+
\frac {16}{\sqrt {27}}
\exp\bigg\{-\frac {k^2}{3\beta_\pi^2}\bigg\}\right] .
\end{equation}
The $I=1$ $\pi\pi$ matrix element from $t$-channel gluon exchange is zero;
this result and the $I=0/I=2$ ratio both
follow immediately from the observation that
$\pi^+\pi^-\not\rightarrow \pi^+\pi^-$ through this mechanism.
Here
$\mu= \cos (\theta_{c.m.})$, where $\theta_{c.m.}$ is the center of
the mass scattering angle, and $k$ is the magnitude of the asymptotic
three-momentum of each meson in the $c.m.$ frame.
This is the matrix element of the spin-spin OGE term
calculated between single-Gaussian $q\bar q$ wavefunctions,
summed over all
four permutations of gluon exchanges and quark line interchanges.
The wavefunction
parameter $\beta_\pi$ is usually taken to be about 0.3 GeV
in the nonrelativistic quark model. In \cite{BS1}
a fit to the $I=2$ $\pi\pi$ S-wave phase shift gave $\beta_\pi = 0.337$ GeV,
$\alpha_s = 0.6$ and $m_q=0.33$ GeV, which we also use here. (Note that the
scattering amplitude (2) from the spin-spin interaction
only involves the combination $\alpha_s/m_q^2$.)

The Hamiltonian matrix element
$h_{fi}$ is
proportional to the Born-order $T$-matrix element
and can be used to calculate Born-order phase shifts
through the relation \cite{BSW}
\begin{equation}\label{8}
\delta_{Born}^{(\ell)}=
-\frac {2\pi^2k E_1 E_2}{{\cal S}E} \int^1_{-1} h_{fi}(\mu)P_\ell(\mu) d\mu \ ,
\end{equation}
where $E_1$ and $E_2$ are the two hadron energies in the c.m. frame
(set equal for $\pi\pi$), $E$ is the total
c.m. energy $E_1+E_2$, and ${\cal S}$ is a statistical factor which is
2 for $\pi\pi$ and 1 for $K\pi$.
The resulting $I=0$ $\pi\pi$ phase shifts for even $\ell$ are
\begin{equation}
\delta_{Born}^{I=0,\ell=even}(t{\rm-ch.}\ {\rm gluon})
=
+\frac {\alpha_s}{9 m_q^2} \, k E_\pi
\left [  \ e^{-x}\; i_\ell (x) +
\frac {8}{\sqrt {27}}\; e^{-4x/3} \;\delta_{\ell,0}
\; \right ]
\end{equation}
where $x =  k^2 / 4\beta_\pi^2$. The S-wave phase shift, using
$i_0(x) = $sinh$(x)/x$, leads to a
scattering length of
\begin{equation}
a_0^{I=0}(t{\rm-ch.}\ {\rm gluon}) = \lim_{k\to 0} \delta^{0,0} / k =
+\frac 19 \bigg( 1 + \frac {8}{\sqrt{27}} \bigg) \frac {\alpha_s m_\pi}
{m_q^2} \ .
\end{equation}
This phase shift and scattering length are positive, corresponding to
an attractive interaction,
but are numerically rather small; with our parameters this
scattering length is
$a_0^{I=0}= +0.043\; {\rm fm}$,
an order of magnitude smaller than
the experimental value of
\begin{equation}
a_0^{I=0}({\rm expt.})= \cases{+0.32(13)\; {\rm fm} \  & (production expts),\cr
                        +0.37(7)\; {\rm fm} \  & ($K_{e4}$ decay). \cr }
\end{equation}
These numbers are taken from a recent review
by Ochs \cite{Ochs}, and
incorporate constraints from dispersion relations.

This conclusion regarding the small contribution of
non-annihilation processes in $I=0$ $\pi\pi$ is more general than this
model, since
$a_0^{I=0} = - \frac 12  a_0^{I=2}$ follows from
$\pi^+\pi^-\not\rightarrow \pi^+\pi^-$
in any single-pair-interchange model, and one
can use the experimental $I=2$ $\pi\pi$ scattering length of
$a_0^{I=2}\approx -0.08\; $ fm
\cite{I2PP}
to normalize the no-annihilation amplitude.
Evidently S-wave $I=0$ $\pi\pi$
scattering is dominated by annihilation processes, and the
interesting question is whether these are due
primarily to a broad $f_0$ $q\bar q$
state (as is often assumed) or
to a broad scalar glueball or another intermediate state such
as $q\bar q g$.

\vskip .5cm
\noindent
2) {\it $s$-channel gluon exchange}

\vskip .5cm

The $s$-channel gluon
exchange scattering mechanism is in many ways the most interesting. We
will first calculate this
contribution using perturbative gluons.
If gluons behaved like photons we could
model the low-energy effects of
$s$-channel gluon exchange
by the familiar positronium annihilation interaction \cite{IZ} augmented by
a color factor,
\begin{equation}
H_I^{pert.}(s{\rm-ch.}\ {\rm gluon}) = +{2\pi\alpha_s\over  m_q^2}\;
(\lambda^a_I/2)\cdot (\lambda^a_F/2)
 \;
 (\vec S_i \cdot \vec S_j+\frac 34 )\  \delta (\vec r_{ij}) \ .
\end{equation}
For $s$-channel transverse gluon exchange this is the leading term in an
expansion in $v^2/c^2$.
There is also a color Coulomb interaction, but we expect this to
be small because the annihilation color charge density
$\langle 0| \rho^a(\vec x \, )
|q\bar q \rangle$ transforms as $L=1$.

A typical annihilation diagram involving this Hamiltonian, for
$K^+\pi^-\to K^+\pi^-$, is shown in Fig.1. The contribution of this diagram
to the meson-meson scattering amplitude can be derived easily using
the diagrammatic techniques discussed in \cite{BS1}, see especially
Appendix C of that reference.
In Ps-Ps scattering this Hamiltonian has a color matrix element
of $+4/9$ and
a spin matrix element of $+3/4$.
We may separate the isospin amplitudes by considering the
special cases $\pi^+\pi^- \to \pi^+\pi^-$
and $\pi^+\pi^- \to \pi^-\pi^+$ ($\hat \Omega_f(\pi^\pm) \to -
\hat \Omega_f(\pi^\pm)$);
this leads to
\begin{equation}
h^{I=0}_{fi}(s{\rm-ch.}\ {\rm gluon}) =
+{2\pi\alpha_s\over  m_q^2}\;
{1\over (2\pi)^3 }\;
\left [
\; e^{
-(1-\mu)x }
+
e^{
-(1+\mu)x }
\; \right ] \
\end{equation}
and
\begin{equation}
h^{I=1}_{fi}(s{\rm-ch.}\ {\rm gluon}) =
+{4\pi\alpha_s\over  3m_q^2}\;
{1\over (2\pi)^3 }\;
\left [
\; e^{
-(1-\mu)x }
-
e^{
-(1+\mu)x }
\; \right ] \ ,
\end{equation}
again using $x =  k^2 / 4\beta_\pi^2$.
The $I=2$ annihilation amplitude is of course zero.
Using (3), these amplitudes lead to $I=0$ and $I=1$
$\pi\pi$ phase shifts of
\begin{equation}
\delta_{Born}^{I,\ell}(s{\rm-ch.}\ {\rm gluon})
=
c_{I,\ell}\, {\alpha_s \over  m_q^2 }\; k E_\pi \; e^{-x}\; i_{\ell}(x)  \ ,
\end{equation}
where $c_{I,\ell}$ is $-1/2$ for $I=0$ and $\ell=$even,
$-1/3$ for $I=1$ and $\ell=$odd, and zero otherwise.
The S-wave phase shift gives an $I=0$ scattering length of
\begin{equation}
a_0^{I=0}(s{\rm-ch.}\ {\rm gluon}) =
-\frac 12  \frac {\alpha_s m_\pi}
{m_q^2} \ .
\end{equation}
This negative phase shift and scattering length correspond to a
repulsive interaction,
opposite to the observed $I=0$ $\pi\pi$ interaction. With our
standard quark model parameters this $s$-channel massless-gluon contribution
is
$a_0^{I=0} = -0.076\; $fm,
somewhat larger than $t$-channel gluon exchange and opposite in sign.

More careful consideration suggests
that the assumption of massless perturbative gluons is
especially unphysical
in this case and should be modified \cite{IS}. One effect of
confinement is to
raise the mass of the lowest intermediate $q\bar q g$ hybrid basis state
from the perturbative value of $2 m_q$ to
the physical hybrid mass $M_H$, which is well above the $\pi\pi$ invariant
masses we consider. This causes the
energy denominator
$1/(E_{q\bar q} - E_g)$ in the $q\bar q\to g \to q\bar q$
subprocess to change sign.
Since the sign of the amplitude for this second-order process
is also
determined by the energy denominator,
confinement changes this sign as well,
which leads
to an attractive force in $I=0$ $\pi\pi$ and $I=1/2$ $K\pi$.
There are many ways one might incorporate this effect of confinement on
virtual gluons; one simple approach is to modify the
denominator of the gluon propagator
by including an effective gluon mass,
\begin{equation}
s^{-1}
\to
\Big( s -\mu_g^2 \Big)^{-1} \ ,
\end{equation}
where the mass scale $\mu_g$ is set by the hybrid mass gap, which is
presumably about 1 GeV.
This modification of the gluon propagator changes the effective low-energy
$q\bar q$ interaction in (7) by replacing $m_q^{-2}$
by $(m_q^2 - \mu_g^2/4 )^{-1}$. This multiplies the naive Hamiltonian
(7) by a
negative constant,
\begin{equation}
H_I^{conft.}(s{\rm-ch.}\ {\rm gluon}) = - \Bigg\{
{ 1\over (\mu_g / 2m_q)^2 - 1 }
\Bigg\}\; \cdot
H_I^{pert.}(s{\rm-ch.}\ {\rm gluon})\ .
\end{equation}
If we estimate constituent masses of
$\mu_g = 1.$ GeV and $m_q = 0.33$ GeV, we expect the $s$-channel
effective interaction to be similar in magnitude to the original
massless gluon form but opposite in sign.
The exact numerical strength of
this diagram is clearly
problematical due to uncertainties in the effect of confinement,
so
in practice we
simply introduce a multiplicative factor $f$ in the $s$-channel
annihilation Hamiltonian and use the form
\begin{equation}\label{2}
H_I^{conft.}(s{\rm-ch.}\ {\rm gluon}) = +f \cdot {2\pi\alpha_s\over  m_q^2}\;
(\lambda^a_I/2)\cdot (\lambda^a_F/2)
 \;
 (\vec S_i \cdot \vec S_j+\frac 34 )\  \delta (\vec r_{ij}) \ .
\end{equation}
The phase shifts (10) and scattering length (11) from $s$-channel gluon
exchange are thus multiplied by a phenomenological parameter $f$.
We will fit
$f$ to the experimental
$I=0$ $\pi\pi$ and $I=1/2$ $K\pi$ S-wave
phase shifts,
with
the caveat that we
expect it to be negative and not far from unity.

\vskip .5cm
\noindent
3) {\it s-channel resonances}
\vskip .5cm

We treat
the phase shift due to
conventional $s$-channel $q\bar q$ resonances
phenomenologically, since it is not clear how to determine
three-meson couplings directly from quark-gluon interactions.
We use a relativized Breit-Wigner form
suggested by the Particle Data Group \cite{PDG1} to model
the $s$-channel resonances,
which has an elastic phase shift of
\begin{equation}\label{12}
\delta_R=\tan^{-1}\bigg\{  \frac {\sqrt{s}\, \Gamma(E)} {
M_R^2-s}\bigg\}\ ,
\end{equation}
where $s=E^2$.
Since we are considering broad resonances it is important to include
the energy dependence of the width $\Gamma(E)$. Of course this
function is somewhat model dependent.
In general we expect it to consist of a
centrifical factor for decay to a
two-body final state with
orbital angular momentum L, times a phase space factor of
$k/M$ and a vertex form factor
$D(k)$ derived from wavefunction
overlaps of the initial and final mesons. Here we use the form
suggested by the LASS collaboration \cite{LASS},
\begin{equation}\label{13}
\Gamma(E) =
\bigg({k \over  k_R}\bigg)^{2L}
\cdot
\bigg({k /  E \over k_R / M_R }\bigg)
\cdot
\bigg({D(k) \over  D(k_R)}\bigg)
\cdot \Gamma(M_R) \ .
\end{equation}
In this formula
$k_R$
is the momentum of one final-state hadron
at the $s$-channel resonance mass,
and for $\pi\pi$,
$k_R=\sqrt{M_R^2/4 -  m_\pi^2}$. (Here and in the remainder of the paper we
assume single decay channels for each resonance, $\pi\pi$ for $f_0$ and $\rho$
and $K\pi$ for $K^*_0$ and $K^*$. The broad scalar resonances couple dominantly
to these channels so this is a reasonable first approximation, especially for
low-energy effects.)

The vertex form factor $D(k)$ is wavefunction-dependent; here we
use
\begin{equation}
D(k) = c \; \exp \bigg\{ - k^2 / 6\beta_\pi^2 \; \bigg \} \ ,
\end{equation}
which follows from the $^3$P$_0$ decay
model with
SHO wavefunctions \cite{LeYaouanc}.
We expect the corresponding form factor for radial excitations to
suppress the low-energy contributions of radially-excited $q\bar q$ states
so they are not important numerically.

The low-energy limit of this resonance phase shift gives the
contribution of a scalar resonance to the scattering length,
which in the $\pi\pi$ (equal mass) case is
\begin{equation}
a_0({\rm res.}) =
{1 \over
(1-4m_\pi^2/M_R^2)^{3/2} }
\ \cdot \ \frac {D(0)}{D(k_R)} \ \cdot {\Gamma(M_R)\over M_R} \
\cdot \ {2 \over M_R} \ .
\end{equation}
For $I=0$ $\pi\pi$ scattering the only well established broad resonance
is the $f_0(1400)$.
The mass and width of this broad state are both rather problematical;
the PDG mass and width estimates are
$M_R\approx 1400$ MeV and
$\Gamma(M_R)\approx 150-400$ MeV.
For this wide range of values the scattering lengths
with the
$^3$P$_0$
form factor are
\begin{equation}
a_0({\rm res.}) = +(0.06-0.17)\; {\rm fm}  \ .
\end{equation}
Taking the recent Crystal Ball central values for the
mass and width from $\gamma\gamma\to \pi^o\pi^o$,
$M=1250$ MeV (assumed) and $\Gamma_R$$ = 268(70)$ MeV
\cite{XB}, we find a
contribution to the scattering length of
\begin{equation}
a_0({\rm res.}) = +0.11(3)\; {\rm fm} \ .
\end{equation}
The Crystal Barrel collaboration \cite{Bugg} recently reported
a similar mass and width for this broad $f_0$,
$M\approx 1335$ MeV and $\Gamma= 255(40)$ MeV, corresponding to a
very similar scattering
length of $0.11(2)\; $fm. The Crystal Barrel and Obelix \cite{Obelix}
collaborations report a
similar, somewhat broader $f_0$
in $4\pi$ final states, with
$M=1374(38)$ MeV and $\Gamma=375(61)$ MeV and
$M=1345(12)$ MeV and $\Gamma=398(26)$ MeV respectively.
These imply a resonance contribution to the scattering length near the upper
limit of (19).

Recall for comparison that the nonresonant gluon exchanges give contributions
to the scattering length of
\begin{equation}
a_0({\rm gluon}\ {\rm ex.})
= \Big\{ +0.043 - 0.076 f \; \Big\} \; {\rm fm} \ ,
\end{equation}
 where $f$ is expected to be negative and not large relative to unity.
Since
the experimental scattering length is about $+(0.3-0.4)\; $fm (6), these
results suggest that
low energy S-wave
$I=0$ $\pi\pi$ scattering
receives important contributions both from the broad $f_0(1400)$
and from nonresonant scattering, and that $f\approx -2$ to $-3$.
Our detailed fits to phase shifts will support similar values for $f$.

The resonance contributions remain large at low energies because
of the
$1/E$ factor in the energy-dependent width $\Gamma(E)$ (16). Although
this appears well motivated as the $1/M$ in the decay rate
of an initial state of mass $M$ \cite{Dunwoodie,BJD},
this factor and the
$\delta(\Gamma,E)$ relation
(15) are so important to the threshold behavior that they
merit more careful study.
It would also be interesting to explore the sensitivity of our
conclusions to the form chosen for
the $k$-dependent three-meson
vertex;
using instead a pointlike-meson form factor, we would predict
a scattering length from the broad $f_0$ resonance about
half as large. This is because
the form factor suppresses the coupling of the
$f_0$ to pions at higher momenta, so if we use the width at
resonance as a fixed input,
the strength of the coupling at threshold is increased by the form factor.
(Note the $D(0)/D(k_R)$ dependence of the
scattering length in (18).)
Another concern
is that
scattering amplitudes from
two different time-orderings,
$\pi\pi\to f_0\to\pi\pi$ and
$\pi\pi\to f_0\pi\pi\pi\pi \to\pi\pi$,
are added with
equal strength to give the covariant form (15). These are
actually modified by
form factors, and the
second ``Z-graph" process may be strongly suppressed
\cite{IMR}. We will test the importance of some of
these complications in our detailed study of
experimental S-wave phase shifts.

\vskip .5cm

\subsection*{3. $I=1/2$ $K\pi$ Scattering Amplitudes}
Application of these techniques to $I=1/2$ $K\pi$ scattering is
straightforward.
First, $t$-channel gluon exchange for $I=1/2$ is simply related to the
$I=3/2$ amplitude,
\begin{equation}\label{20}
h^{I=1/2}_{fi}(t{\rm-ch.}\ {\rm gluon})=
-{1\over 2} \, h^{I=3/2}_{fi}(t{\rm-ch.}\ {\rm gluon}) \ ,
\end{equation}
which follows from
$K^-\pi^+ \not\rightarrow
K^-\pi^+ $ through this mechanism.
Taking the $I=3/2$ result from \cite{BSW}, we have
\begin{equation}\label{21}
h^{I=1/2}_{fi}(t{\rm-ch.} \ {\rm gluon})=
-\frac 1{(2\pi)^3}\frac {2\pi\alpha_s}{9m_q^2}\left (T_1+
T_2+C_1+C_2\right ),
\end{equation}
where $T_i$ and $C_i$ represent the contributions of
the
``transfer" and ``capture" diagrams \cite{BS1}.
Specializing to the case of identical
pion and kaon spatial wavefunctions ($\beta_\pi = \beta_K$), these
are explicitly
\begin{equation}\label{22}
T_1=\exp
\bigg\{ -{ (2+2\zeta+\zeta^2)
(1-\mu)\over 2}\, x
\bigg\},
\end{equation}
\begin{equation}\label{23}
T_2=\rho \ \exp \bigg\{-{
 2 - 2\zeta + \zeta^2+2(1-\zeta)\mu\over 2}\, x
\bigg\},
\end{equation}
\begin{equation}\label{24}
C_1=\rho
 \ (4/3)^{3/2} \exp \bigg\{-{
4-\zeta+\zeta^2 -3\zeta\mu \over 3}\, x
\bigg\}
\end{equation}
and
\begin{equation}\label{25}
C_2= (4/3 )^{3/2} \exp \bigg\{-
{4+\zeta+2\zeta^2
-(5\zeta + \zeta^2)\mu\over 3}\, x\bigg\} ,
\end{equation}
where $x= k^2/ 4 \beta_\pi^2$, as in $\pi\pi$ scattering.
These somewhat complicated results allow for different strange
and nonstrange quark masses
through the parameter
\begin{equation}\label{26}
\rho=m_q/m_s  \ ,
\end{equation}
and $\zeta$ is the combination $(1-\rho)/(1+\rho)$.
In our previous study of $I=3/2$ $K\pi$ scattering
\cite{BSW} we considered the more general problem of
different pion and kaon length scales, but found that the
$I=3/2$ scattering amplitudes
were quite insensitive to the ratio $\beta_\pi/\beta_K$, and setting it
equal to unity
allowed a very good description of the data. For
equal length scales we found an optimum value
of $\rho = 0.677$, which we will also assume here.
\vskip 0.2cm

The $s$-channel gluon exchange diagram
with $\beta_\pi=\beta_K$ gives a Hamiltonian matrix element of
\begin{equation}
h^{I=1/2}_{fi}(s{\rm-ch.}\ {\rm gluon})=
+f\cdot
\frac {\pi\alpha_s}{m_q^2}
\frac 1{(2\pi)^3}
\exp
\bigg\{ -{
(2+ 2\zeta + \zeta^2 )(1-\mu)\over 2}\, x
\bigg\},
\end{equation}
which by inspection is proportional to the contribution of the $t$-channel
gluon diagram ``transfer$_1$" ($T_1$ above), and so can be incorporated
in (23) by the substitution $T_1\to (1-9f/2) T_1$.
The general-$\ell$ phase shift  and the corresponding scattering length
from combined $s$- and $t$-channel gluon exchange are
\begin{displaymath}
\delta^{I=1/2,\ell}({\rm gluon}\ {\rm ex.}) =
{\alpha_s \over 9 m_q^2}\; {k E_\pi E_K \over E}
\bigg[
\ (1-\frac 92 f)\, e^{-(1+\zeta + \frac 12 \zeta^2)x}
\ i_\ell\Big(
{2+2\zeta + \zeta^2\over 2}\, x
\Big)
\end{displaymath}
\begin{equation}
+ \rho \ \Big\{ e^{-(2-\zeta + \frac 12 \zeta^2)x}
+
\,  (4/3)^{3/2} \; e^{-{1\over 3}(4-\zeta + \zeta^2)x}
\; \Big\} \; i_\ell(
\zeta x
) +
(4/3)^{3/2}\;  e^{-{1\over 3}(4+\zeta + 2\zeta^2)x}
\ i_\ell\Big(
{5\zeta + \zeta^2\over 3}\, x
\Big)
\bigg]
\end{equation}
and
\begin{equation}
a_0^{I=1/2}({\rm gluon} \ {\rm ex.}) = {\alpha_s \over 9 m_q^2} \;
{m_\pi m_K\over m_\pi + m_K}
\left [
(1 + (4/3)^{3/2} )
(1+\rho )
- \frac 92 f \; \right ] \ .
\end{equation}

The scalar
$K_0^*(1430)$ is
the single important $s$-channel
resonance in low-energy S-wave $I=1/2$ $K\pi$ scattering.
We incorporate
this state using a relativized Breit-Wigner of the
form (15-17), as in $\pi\pi$ scattering. It leads to a scattering length of
\begin{equation}
a_0^{I=1/2}({\rm res.}) = {M_R^2\over  M_R^2
 - (m_\pi + m_K )^2 } \
{M_R\over k_R} \
{D(0) \over D(k_R)} \ {\Gamma(M_R) \over M_R} \ {1\over M_R} .
\end{equation}
The PDG mass and width for this state (taken from LASS results) are
$M=1429\pm 4\pm 5$ MeV and
$\Gamma=287\pm 10\pm 21$ MeV \cite{PDG}, but subsequent reanalysis
by this group has led
to preferred values of
$M=1412$ MeV and
$\Gamma=294$ MeV \cite{Dunwoodie}.
These correspond
to a contribution to the scattering length of
$a_0(K^*_0(1430)) = 0.145 \ $fm given our form for the relativized Breit
Wigner (presumably with a statistical error from $\Gamma$ of about $10 \% $).
In summary, the total scattering length we expect from gluon exchange and
the $s$-channel resonance $K^*_0(1430)$ is numerically (excluding errors)
\begin{equation}
a_0^{I=1/2}(K\pi \ {\rm thy.}) = \bigg\{\; 0.056 \;
-  0.059 f \; + 0.145 \bigg\} \ {\rm fm} \ .
\end{equation}
which we may compare with the experimental scattering length
found by Estabrooks {\it et al.} \cite{Estakp},
\begin{equation}
a_0^{I=1/2}(K\pi \ {\rm expt.}) = 0.472(8) \ {\rm fm} \ .
\end{equation}
Evidently the scattering lengths (33) and (34)
suggest a value of $f\approx -4$ for the
$s$-channel gluon strength parameter, although the
individual contributions to the scattering length
have important uncertainties so this parameter is not very
well determined. In our subsequent analysis of phase shifts we shall
find
that
a somewhat smaller value of
$f\approx -2.5$ appears reasonable over a wide range of energies for
$\pi\pi$ and $K\pi$ elastic scattering.

\subsection*{\bf 4. Comparison to Experimental Phase Shifts}

\vskip .5cm
\noindent
{\it a) $I=0$ $\pi\pi$ and $I=1/2$ $K\pi$ S-waves}
\vskip .5cm

We begin our detailed comparison with experiment with a study of the
$I=0$ $\pi\pi$ and $I=1/2$ $K\pi$
S-waves.
We combine these because we expect them to be closely related by
SU(6) symmetry,
and together they provide data on
nearly elastic S-wave Ps-Ps scattering amplitudes
from
invariant masses of $\approx 0.3 $ to 1.6 GeV.

We impose unitarity on the total scattering amplitude by assuming
that the phase shifts for the individual processes add,
as they would for small amplitudes;
\begin{equation}
\delta_{tot}=\delta_{Born}+\delta_R\ .
\end{equation}
This is equivalent to the
prescription used by the LASS collaboration, which was to
assume a relative phase between
resonant and background amplitudes, chosen so that
the complex sum remains on the
unitarity circle,

\begin{equation}
\sin (\delta_{tot})
e^{i\delta_{tot}}
=
\big[
\sin (\delta_{bkg})
e^{i\delta_{bkg}}
\big]
+ e^{2i\delta_{bkg}} \cdot
\big[
\sin (\delta_R)
e^{i\delta_R}
\big]\ .
\end{equation}

For these amplitudes the Born phase shift is given by (30) and the
$f_0(1400)$ and $K^*_0(1430)$ resonance
phase shifts by (15-17).
We fix the quark model parameters at $\alpha_s = 0.6$, $m_q=0.33$ GeV,
$\rho=0.677$ and $\beta_\pi=\beta_K$ (from our previous study of
$I=2$ $\pi\pi$ \cite{BS1} and $I=3/2$
$K\pi$ \cite{BSW} phase shifts).
We generally set the pion and kaon masses equal to the
isospin averages
$m_\pi = 0.138$ GeV
and
$m_K = 0.495$ GeV, but for the special case $I=0$ $\pi\pi$
we use
$m_\pi = m_{\pi^+} = 0.1396$ GeV, as appropriate for the near-threshold
points from
$K_{e4}$ decay.

The
$I=0$ $\pi\pi$ channel is the most interesting historically
\cite{PPHist},
due to long-standing
uncertainties in the properties of the very broad $f_0$ resonance
seen
in this slowly rising phase shift.
This channel is further complicated by
inelasticities associated with the narrow $f_0(975)$,
and a scalar glueball is expected near 1.5 GeV, so in
this single channel study we consider only invariant masses below 0.9 GeV.
For our $I=0$ $\pi\pi$
data set we use the $s$-channel and $t$-channel extrapolations of
Estabrooks and Martin \cite{EM} (38 points from 0.51 GeV to 0.89 GeV),
the ``case 1" phase shift of Protopopescu {\it et al.} \cite{Proto}
(17 points from 0.55 GeV to 0.89 GeV), and the low-energy data
from $K_{e4}$ decay of Rosselet {\it et al.} \cite{Ross} (5 points for
$\delta^{I=0,\ell=0}-
\delta^{I=1,\ell=1}$ from 0.289 GeV to 0.367 GeV). We added an estimated
low-energy P-wave phase shift proportional to $k_\pi^3$ to the Rosselet
data, with a coefficient chosen to give $9.4^o$ at 0.51 GeV, as reported by
Estabrooks and Martin \cite{EM}.

For our $K\pi$ data set we use
the 24 data points of Estabrooks {\it et al.} \cite{Estakp}
(from 0.73 to 1.30 GeV, the full set of separated-isospin phases quoted)
and the 37 data points of Aston {\it et al.} \cite{LASS} below 1.6 GeV.
This cutoff was chosen because the radial $K_0^*(1950)$
should become important above this mass. Since the LASS collaboration
\cite{LASS} tabulated
only the $K^-\pi^+$ phase shift $\phi$
\cite{Awaji} rather than the separated
isospin ones, we used our formalism to calculate $I=3/2$ S-wave shifts
as well to
fit this $\phi$ data directly.
We will show the results of the fit for $I=1/2$
phase shifts, and the experimental errors we show in the figure
are actually those quoted by LASS for the
mixed-isospin phase $\phi$. This gives a total data set of 121 points
for S-wave Ps-Ps scattering, from
an invariant mass of 0.289 GeV in $\pi\pi$ to 1.59 GeV in $K\pi$.

We fitted this full S-wave data set with the single $s$-channel annihilation
strength parameter $f$ and masses and widths for the two broad resonances
$f_0(1400)$ and $K^*_0(1430)$. Since the mass of the
$f_0(1400)$ is poorly determined due to our
0.9 GeV cutoff in the $\pi\pi$ data,
we constrained the masses by $M_{f_0} = M_{K^*_0} - 0.12$ GeV,
as suggested by the masses of other members of the $^3$P$_{\rm J}$ SU(3)
flavor
multiplet. There is some indication of an $f_0(975)$ contribution to the
phase shift near 0.9 GeV, so we also added a conventional elastic Breit-Wigner
phase shift with $M=0.974$ MeV and $\Gamma$=0.047 Gev (PDG values, both fixed)
to the
$I=0$ $\pi\pi$
phase shift.
The optimum values of the parameters were found to be $f=-2.573$,
$M_{K^*_0} = 1.477$ GeV,
$\Gamma_{K^*_0} = 0.261$ GeV,
$M_{f_0} = 1.357$ GeV (constrained by $M_{K^*_0}$), and
$\Gamma_{f_0} = 0.405$ GeV.
The experimental and fitted phase shifts are shown in Fig.2 ($K\pi$) and
Fig.3 ($\pi\pi$), and except for some inaccuracies in describing the
shape of the $K^*_0(1430)$ resonance region these
four parameters evidently give a reasonable
description of both amplitudes.

The resonance widths found in this fit
are especially attractive;
$\Gamma_{f_0} = 0.405$ GeV is reasonably consistent with the recent
experimental values of
$0.255(40)$ \cite{Bugg},
$0.268(70)$ \cite{XB},
$0.375(61)$ \cite{Bugg}, and
$0.398(26)$ GeV \cite{Obelix}.
The discrepancy between these moderate widths (from production
of the $f_0$ in $\gamma\gamma$ and $P\bar P$)
and the very large widths seen in scattering can thus
be understood as due to the additional nonresonant contributions to scattering,
which make the $f_0$ and $K^*_0$
appear broader than they actually are.
It is also reassuring that the relative fitted
$f_0$ and $K^*_0$
widths are not far from the naive SU(6) expectation of
$\Gamma_{f_0}/\Gamma_{K^*_0} = 2$ (assuming only $\pi\pi$ and $K\pi$ modes
and neglecting phase space differences).

The fitted
mass for the broad $f_0$, $M_{f_0} = 1.357$ GeV,
compares well with experimental values
from recent $P\bar P$ annihilation experiments,
$1.335(30)$ \cite{Bugg},
$1.345(12)$ \cite{Obelix}, and
$1.374(38)$ GeV \cite{Bugg}.
Our fitted $K^*_0$ mass of 1.477 GeV is significantly
higher than the LASS value of
1.412 GeV, although an independent analysis of the LASS data by Weinstein and
Isgur \cite{WKPS} using a two-channel
Schr\"odinger formalism found 1.47 GeV, essentially
identical to our result. The discrepancy appears to be due
mainly to the use by
LASS of the phase shift formula (15) with $\sqrt{s}$ replaced by $M_R$; on
fitting our $K\pi$ S-wave data set with their form we find
$M_{K^*_0} = 1.432$ GeV. Although the most reasonable generalization of the
Breit-Wigner form to a broad resonance
is not well established theoretically, we found that
(15) with $\sqrt{s}$ replaced by $M_R$
gives an unsatisfactory fit to the low energy $I=0$ $\pi\pi$ data
(a shoulder is predicted near threshold), and for this reason
we have used the PDG form (15).

The scattering lengths in this fit and the quark Born
diagram (nonresonant) and resonance
contributions are as follows:
\begin{equation}
a_0^{I=0}(\pi\pi) =
0.41 \ {\rm fm}\ =
0.24 \ {\rm fm (nonres.)} +
0.17 \ {\rm fm (res.)} \ ,
\end{equation}
\begin{equation}
a_0^{I=1/2}(K\pi) =
0.33 \ {\rm fm}\ =
0.21 \ {\rm fm (nonres.)} +
0.12 \ {\rm fm (res.)} \ .
\end{equation}
Although Pennington and especially Burkhardt and Lowe \cite{Ochs}
prefer
a value closer to 0.3 fm for the
$I=0$ $\pi\pi$ scattering length \cite{Ochs},
it is clear from Fig.3 that our fit (with 0.41 fm) gives
a satisfactory description of the
existing low-energy $I=0$ $\pi\pi$ data, and considerably
improved low energy
measurements will be required to distinguish these values.
The larger value of about 0.472(8) fm
cited by Estabrooks {\it et al.} \cite{Estakp}
for the $I=1/2$ $K\pi$ scattering
length
may give a somewhat
improved fit to the low-energy $K\pi$ data
(see Fig.2),
but a better measurement of
S-wave phase shifts near and below 0.9 GeV
would be required to confirm this larger value.
If improvements in the low energy data do confirm a value of
0.3 fm for $I=0$ $\pi\pi$, this
could be accommodated in our
formalism by reducing the broad $f_0$ width to about 300 MeV and changing
$f$ to $-2.$ It is difficult however
to see how a $K\pi$ scattering length of
0.47 fm could be fitted simultaneously without unrealistic
parameter changes. For this reason we suggest that
the error in (34) was underestimated, and
that a value near 0.3 fm will be found in a higher-statistics
measurement of low energy $I=1/2$ $K\pi$ scattering.

To test the stability of the fitted parameter
values we performed fits with $f$ fixed and the resonance
masses and widths free (but with $M_{K^*_0} - M_{f_0} =0.12$ GeV
constrained). The variation of the
fit residual
$F = \kappa \sum_i (\delta_i({\rm expt.})
- \delta_i({\rm thy.}))^2/\epsilon_i({\rm expt.})^2$
and the resonance parameters with $f$ is shown in Fig.4. (We normalize
$F$ to unity for $f=0$.)
Evidently nonzero $f$ improves the fit considerably, by better than
a factor of two relative to $f=0$. There is a strong correlation between $f$
(which provides a smoothly rising phase shift) and the resonance widths;
with $f=0$ the best fit requires implausibly large widths of
$\Gamma_{f_0} = 1.47$ GeV and
$\Gamma_{K^*_0} = 0.57$ GeV.
As the nonresonant scattering is increased with
$-f$, the widths of the resonances required to describe the balance of the
low-energy
scattering fall. By $f=-2$ the widths have fallen to the
more reasonable values
$\Gamma_{f_0} =   0.61$ GeV and
$\Gamma_{K^*_0} = 0.34$ GeV. There is a region of comparable quality of fit,
$-2 > f > -3$, although by $f=-3$ the fitted $f_0$ width is
$0.27$ GeV, near the lowest experimental estimates.

Finally, we tested some of the uncertainties in modelling broad resonances
discussed in Sec.2.3 by carrying out fits to the $K\pi$ data set
alone with
different forms for the resonance. The PDG form (15)
gives
$(M_{K^*_0}(GeV), \Gamma_{K^*_0}(GeV), f) = $
$(1.477, 0.257, -2.602)$;
the LASS form (as discussed above) gives
$(1.432, 0.308, -1.403)$; the LASS form with a single time ordering
(which is a nonrelativistic Breit-Wigner with the energy-dependent
width (16)) gives
$(1.457, 0.284, -2.011)$;
the LASS form with no meson form factor gives
$(1.477, 0.268, -2.584)$; and finally, a simple nonrelativistic Breit-Wigner
phase shift with no meson form factor and a constant width gives
$(1.471, 0.290, -2.393)$. Evidently the various assumptions about the resonance
phase shift lead to variations of about $\pm 20$ MeV in the mass and width
and about
$\pm 0.5$ in the parameter $f$.

\vskip .5cm
\noindent
{\it b) $I=1$ $\pi\pi$ and $I=1/2$ $K\pi$ P-waves}
\vskip .5cm

The higher partial waves may be treated similarly, and there are fewer
complications because the resonances are narrower and the
Born contributions are smaller. Indeed, a casual inspection of the
$I=1$ $\pi\pi$ and
$I=1/2$ $K\pi$
P-wave phase shift data to 1.5 GeV shows clear evidence for
the $\rho(770)$ and $K^*(892)$ and little else.

For our $I=1$ $\pi\pi$ P-wave
data set we again use the $s$-channel and $t$-channel extrapolations of
Estabrooks and Martin \cite{EM} (42 points from 0.51 GeV to 0.97 GeV) and
the ``case 1" phase shift of Protopopescu {\it et al.} \cite{Proto}
(26 points from 0.55 GeV to 1.15 GeV).
For $K\pi$ we choose to fit only the Estabrooks et al.
\cite{Estakp} P-wave data (24 points, at energies
described above); since the $I=3/2$
P-wave is not well established we cannot use the
mixed-isospin P-wave phase shift
$\phi$ tabulated in the LASS data \cite{Awaji}.

To fit these
P-waves we again use the resonant phase shift (15-17)
and set $\ell=1$ in (30).
The quark model parameters $\alpha_s$, $m_q$, $m_s$ and $\beta_\pi = \beta_K$
and the meson masses $m_\pi $ and $m_K$
were assigned the same values as in the S-wave fit.
Since the quark Born diagram contributions to the P-wave phase shifts are
rather small (typically about 5$^o$ at 1 GeV invariant mass), we do not
fit them independently but instead assume the value of $f$
found in the S-wave scattering amplitudes, $f=-2.573$. This leaves
the $\rho$ and $K^*$ masses and widths as free parameters. Fitting the
92 P-wave phase shift data points gives
$M_\rho = 0.7703$ GeV, $\Gamma_\rho = 0.1563$ GeV,
$M_{K^*} = 0.8950$ GeV and $\Gamma_{K^*} = 0.0544$ GeV; these
are in quite good agreement with PDG values, the largest discrepancies being
about 5 MeV in both widths. Note that the nonresonant amplitudes play an
important part in bringing about this close agreement; imposing $f=0$ makes
the fit less than half as accurate (it more than doubles $F$) and leads to
$\rho$ parameters of
$M_{\rho} = 0.7633$ GeV and $\Gamma_\rho = 0.1434$ GeV, which do not reproduce
the
PDG values of
$M_\rho = 0.7681(5)$ GeV and $\Gamma_\rho = 0.1515(12)$ GeV so well.

The general features of the $\pi\pi$ and $K\pi$ fits are quite similar, so
we display results from $K\pi$ only, which has smaller errors at the
higher energies. The fit to the $I=1/2$ $K\pi$ P-wave is shown in Fig.5, which
also shows the individual contributions from the $K^*(892)$ and the
$s$- and $t$-channel Born diagrams. Evidently the quark Born diagram
contribution is dominated by $s$-channel gluon exchange (hybrid diagrams).
The $t$-channel diagrams give zero for $I=1$ $\pi\pi$ scattering since they
are even under $\theta\to -\theta$. The $K\pi$ P-wave is related to the
$\pi\pi$ P-wave by SU(6) symmetry, and the
$K\pi$ $t$-channel contribution is
nonzero only because of SU(6) violation through $m_s\neq m_{u,d}$.
It appears that the resonance shape alone does not give a good description
of the phase shift somewhat above the $K^*$ (note especially the $0.95$ to
$1.25$ GeV region,
and the Born diagrams (with no free parameters) give just
about the contribution required for
good agreement with experiment.
This may be correct, but we must be cautious in this interpretation; the
smaller resonance contribution is due to the energy-dependent width
$\Gamma(E)$ in (15), which is increasing quite rapidly for the $K^*$ in
this mass region.
A similar effect is seen in the $I=1$ $\pi\pi$ fit.
Since this $\Gamma(E)$ is rather uncertain, we can only
conclude that the quark Born diagrams give a contribution to P-wave
phase shifts which is consistent with data, and may be observable somewhat
above the $\rho$ and $K^*$ masses. The systematic uncertainties
in describing the resonance phase shifts however are comparable to the
Born diagram contributions, and until the resonance contributions
can be established with better accuracy we
cannot claim to have confirmed the predicted
quark Born amplitudes in this P-wave data.

\subsection*{\bf 5. Conclusion and Suggestions for Future Studies}

We have extended the constituent interchange model of Barnes and Swanson
to processes with valence $q\bar q$ annihilation by incorporating
$s$-channel gluon exchange and $s$-channel relativized Breit-Wigner resonances.
We applied these techniques to $I=0$ and $I=1$ $\pi\pi$ and $I=1/2$ $K\pi$
scattering in S- and P-waves, since these amplitudes are well established
experimentally and can be studied for evidence of
nonresonant scattering in addition to
the known $s$-channel resonances. In a simultaneous fit to the $\pi\pi$ and
$K\pi$ S-waves we determined the strength of the $s$-channel gluon exchange
(hybrid) diagram and fitted the masses and widths of the $K^*(1430)$ and the
problematical $f_0(1400)$. In our best fit we find an $f_0$ mass and width
of 1357 MeV and 405 MeV, comparable to recent results from $\gamma\gamma$ and
$P\bar P$ production experiments. We conclude that most of the low energy
S-wave scattering in these channels is due to nonresonant $s$-channel
gluon annihilation, which makes the resonances appear very broad in phase shift
data. We also applied these techniques to the $I=1$ $\pi\pi$ and $I=1/2$
$K\pi$ P-wave phase shifts, and concluded that the $s$-channel resonances
$\rho(770)$ and $K^*(892)$ alone suffice to describe most of the P-wave
phase shifts from threshold to over 1 GeV. There may be evidence for
the rather small contribution expected from the quark Born diagrams
in the P-waves near and above 1 GeV.

We repeatedly found that uncertainties in the generalization
of the nonrelativistic Breit-Wigner phase shift to broad resonances
limited our ability to separate the resonant and nonresonant contributions to
scattering in these channels.
For this reason it would be very useful
to establish the limitations and range of validity
of the PDG form (15) we have assumed in most of this work.

These
techniques
could be applied most usefully to reactions
in which nonresonant
$s$-channel gluon exchange dominates
$s$-channel $q\bar q$ resonances.
One such possibility
is $P\bar P\to \Lambda
\bar \Lambda$ \cite{LEAR},
which has been the subject
of detailed experimental investigation at LEAR.
It would be
interesting to determine whether the $s$-channel gluon Hamiltonian (14), with
strength $f$ fitted to Ps-Ps elastic scattering,
also gives an accurate
description of this $q\bar q \to s\bar s$ annihilation process. This study
would presumably require incorporation of
initial- and final-state interactions in a coupled channel formalism.
This rection has already attracted considerable theoretical interest, and
analyses using similar techniques have been reported \cite{KW}.

\subsection*{\bf Acknowledgment}
We would like to thank D.Aston, D.V.Bugg, S.Capstick,
F.E.Close, W.Dunwoodie, D.Herzog, N.Isgur, L.Kisslinger,
C.Meyers, D.Morgan, C.Morningstar, M.R.Pennington,
B.N.Ratcliff, S.R.Sharpe and J.Weinstein for discussions of this material,
and D.Aston for providing unpublished LASS data used here.
This work was supported in part by
National Science Foundation grant PHY-9023586 at Carnegie-Mellon
University and by the United States Department of Energy under contracts
DE-AC02-76ER03069 with the Center for Theoretical Physics at the Massachusetts
Institute of Technology and DE-AC05-840R21400
at Oak Ridge National Laboratory
with Martin Marietta Energy
Systems, Inc.
ZL would like to
thank the Theoretical Physics division of
Oak Ridge National Laboratory
for their hospitality in the course
of this work.

\end{document}